\documentclass[fleqn,usenatbib]{mnras}

\usepackage{newtxtext,newtxmath}

\usepackage[T1]{fontenc}
\usepackage{ulem}

\DeclareRobustCommand{\VAN}[3]{#2}
\let\VANthebibliography\thebibliography
\def\thebibliography{\DeclareRobustCommand{\VAN}[3]{##3}\VANthebibliography}

\usepackage{graphicx}	
\usepackage{amsmath}	

\title[BDs in 47\,Tuc]{\textit{JWST} unveils the brown dwarf sequence of 47\,Tucan\ae}

\author[D. Nardiello et al.]{
D.\ Nardiello$^{1,2}$\thanks{E-mail: domenico.nardiello@inaf.it},
M.\ Griggio$^{1,3}$,
and L.\ R.\ Bedin$^{1}$ \\
$^{1}$Istituto Nazionale di Astrofisica, Osservatorio Astronomico di Padova, Vicolo dell'Osservatorio 5, Padova I-35122, Italy\\
$^{2}$Aix Marseille Univ, CNRS, CNES, LAM, Marseille, France \\
$^{3}$Dipartimento di Fisica, Universit\`a di Ferrara, Via Giuseppe Saragat 1, Ferrara I-44122, Italy\\
}

\date{Accepted 2023 February 9. Received 2023 February 8; in original form 2023 January 24}

\pubyear{2023}

\begin{document}
\label{firstpage}
\pagerange{\pageref{firstpage}--\pageref{lastpage}}
\maketitle

\begin{abstract}
We have developed a technique to restore scientific usage in
compromised (publicly-available) images collected with the
\textit{James Webb Space Telescope} ({\it JWST}) of the Galactic globular
cluster NGC\,104 (47\,Tucan\ae).
In spite of the degradation and limited data, we were able to recover
photometry and astrometry for the coolest stellar objects ever
observed within a globular cluster, possibly unveiling the brightest
part of the brown dwarf (BD) sequence.
This is supported by: 
\textit{(i)} proper motion membership, derived by the comparison with
positions obtained from \textit{Hubble Space Telescope} archival early
epochs; \textit{(ii)} the predicted location of the BD sequence; and
\textit{(iii)} the mass function for low-mass stars derived from models.
Future \textit{JWST} observations will provide the necessary deep and
precise proper motions to confirm the nature of the here-identified BD
candidates belonging to this globular cluster.
\end{abstract}




\section{Introduction}

\begin{figure*}
\includegraphics[bb=33 322 549 698, width=0.95\textwidth]{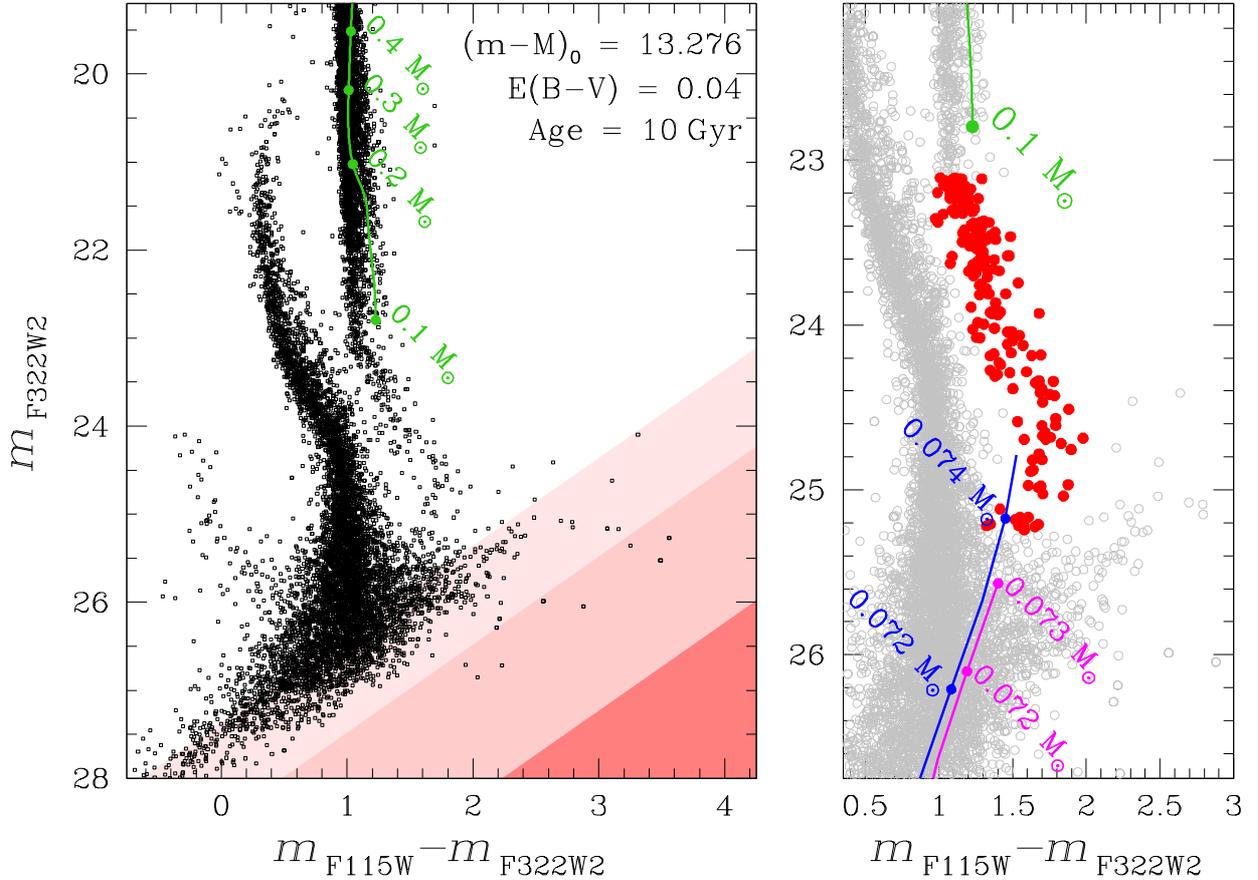}
\caption{Left panel: $m_{\rm F322W2}$ versus $m_{\rm F115W}-m_{\rm
    F322W2}$ CMD for all the well measured stars in the {\it JWST}
  field. Green line represents the 10~Gyr \texttt{BaSTI-IAC}
  isochrone. Shaded regions are our detection limits in both the
  filters: dark red corresponds to the background noise ($1\sigma$)
  converted in magnitude, middle and light red are the $3\sigma$ and
  $5\sigma$ limits, respectively. Right panel is a zoom of the CMD
  around the region where the high-mass BDs are expected. We
  highlighted in red objects whose mass should be between 0.07 and
  0.1~$M_{\odot}$. In blue and magenta are reported the
  solar-metallicity BD 10~Gyr isochrones (ATMO2020 CEQ and NEQ strong models,
  respectively) from \citet{2020A&A...637A..38P}. \label{fig:0}}
\end{figure*}

Globular clusters (GCs) have always been fundamental benchmarks to
test models of stellar evolution, as they are formed by stars
spanning a wide range of masses and with roughly the same age, chemical
composition and distance. GCs are among the oldest known objects of
the Milky Way and possibly in the Universe, and their properties can
be determined using stellar colour-magnitude diagrams (CMDs).

The {\it Hubble Space Telescope} ({\it HST}) has made significant
contributions to our understanding of these systems, both in the
Galactic (e.g., \citealt{2004ApJ...605L.125B,2007ApJ...661L..53P})
and extra-galactic context (\citealt{2019MNRAS.485.3076N}). With its
superior resolution and infrared (IR) sensitivity we can expect that
the {\it James Webb Space Telescope} ({\it JWST}) will also greatly
enhance the study of GCs.
Arguably, the most probable finding of {\it JWST} will be the complete
characterisation of GCs down to the entire lowest masses along the
Main Sequence (MS), and in the so far unexplored regions of 
brown dwarfs (BDs).

BDs are objects that do not have enough mass to fuse Hydrogen in their
cores and instead contract to the size of Jupiter, cooling and dimming
over time (\citealt{1963PThPh..30..460H,1963ApJ...137.1121K}). While
thousands of BDs of Solar metallicity have been identified and studied
near the Sun and in nearby young clusters and associations
(\citealt{1995Natur.377..129R,
  2019ApJS..240...19K,2021AJ....161...42B}), they have not yet been
observed in GCs despite dedicated searches
(\citealt{2016ApJ...817...48D,2019MNRAS.486.2254D}). Identifying BDs
in old and metal-poor GCs would greatly improve our knowledge of
BDs and GCs. Indeed, the chemical composition of BDs has a major
impact on their properties.  For example, the mass at the Hydrogen
burning limit (HBL) is set by the internal opacity and metallicity:
metal-free BDs have higher HBL mass (0.090\,$M_\odot$) than BDs of Solar
abundance (0.072\,$M_\odot$; \citealt{1994ApJ...424..333S}).
Cooling is also affected by the chemical composition through
opacity. Low temperatures allow for the formation of complex molecules
in BDs' atmospheres, which in turn, affect significantly their spectra
(\citealt{2020ApJ...892...31B}).
As BDs cool down (i.e., age), they increasingly separate in luminosity from
stars. In the case of the extreme ages of GCs ($>$10\,Gyrs), a
significant \textit{gap} in luminosity is expected, the width of which
depends on the interior equation of state and low-temperature fusion
processes.
Studies of GCs with the \textit{HST} have identified the beginning of
this gap, but the most massive BDs (the brightest) at the bottom of
the gap have not been detected yet. It is expected that the
capabilities of {\it JWST} will reach these BDs in GCs, allowing for
an independent measurement of the age of GC populations through the
HBL gap (\citealt{2017arXiv170200091C,2019BAAS...51c.521C}).

Concerning the multiple population phenomenon observed within GCs
(\citealt{2015AJ....149...91P, 2015MNRAS.451..312N,
  2018MNRAS.477.2004N, 2019MNRAS.489L..80B}), particularly in the
most massive and dynamically less evolved ones
(\citealt{2004ApJ...605L.125B,2007ApJ...661L..53P,2007ApJ...667L..57S}),
BDs can provide important information. 
Indeed, BDs are highly sensitive to variations in composition
and can serve as amplifiers of these differences, which can help
to fine-tune models of the atmospheres of ultra-cool stars and
sub-stellar objects of different masses, ages, and chemical
properties. This can help relate the properties of observed stellar
populations at higher masses, in order to better understand the
observed differences in chemical compositions among multiple
generations of stars, which are not yet fully understood
(\citealt{2015MNRAS.454.4197R}).

In this letter, we derived photometry and astrometry for the
coolest stellar objects ever observed in a GC, possibly unveiling for
the first time the brightest part of the BD sequence in a GC.

\begin{figure*}
\includegraphics[bb=22 147 551 698, width=0.7\textwidth]{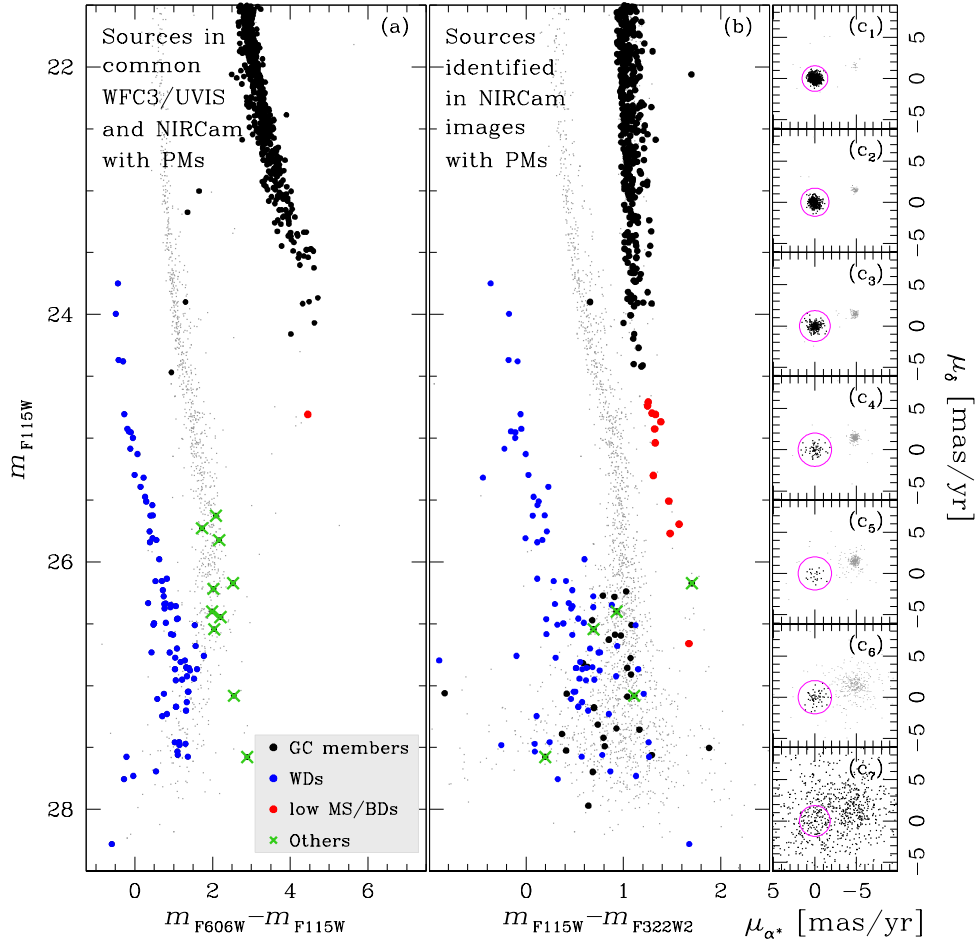}
\caption{Identification of 47\,Tuc's members through PMs. Panel (a)
  shows the $m_{\rm F115W}$ versus $(m_{\rm F606W}-m_{\rm F115W})$
  CMD, used to identify likely WDs (in blue) and faint objects visible
  in the F606W band (green crosses). These selections are used to
  exclude some candidate low MS/BDs (in red) in the $m_{\rm F115W}$
  versus $(m_{\rm F115W}-m_{\rm F322W2})$ CMD of panel (b), that shows
  all the stars identified in the {\it JWST} data and for which we
  measured PMs with F606W, F110W, and F160W images. Panels (c)
  illustrate the PMs in the magnitude intervals corresponding to the
  ordinate axis of panels (a) and (b); black stars in the magenta
  circles are the likely GC members. \label{fig:1}}
\end{figure*}

\section{Observations and data reduction}
%
\label{sec:obs}

The data employed in this paper are \textit{JWST} public archival
material which was compromised by technical problems. The program is
GO-2560 (PI: Marino) which focus on characterizing multiple stellar
populations in very-low mass M-dwarfs of the relatively close and
massive Galactic GC 47\,Tucan\ae\, (NGC\,104).
Visit 13 of GO-2560 suffered of a guide star acquisition failure, that
has resulted in imperfect images. These images were judged unsuitable
for the main science of that program and requested (and approved) to
be repeated in June 2023.
The discarded data, according to STScI policies, becomes immediately
public.

The data for this article was collected using the Near Infrared Camera
(NIRCam) on {\it JWST} and centred on a region located about 6~arcmin
from the 47\,Tuc's centre. The observations took place on July 13--14,
2022 and utilised two filters: F115W (short wavelength) and F322W2
(long wavelength); 40 exposures of 1030.7s were taken with each
filter, using the \texttt{DEEP8} readout mode. 
The observations were collected with two large dithers to fill gaps between the
detectors within a module, but not the gap between the two modules.

In our efforts to extract optimal astrometric and photometric
capabilities of \textit{JWST} cameras
(\citealt{2022MNRAS.517..484N,2022arXiv221203256G}), we searched the
archive for dense stellar fields, and quickly realised that these
failed images are an optimal benchmark to test our algorithms.  Our
procedures derive completely empirical effective PSFs (ePSFs) of any
shape, and rely on well-characterised geometric distortion for the
cameras. In the following, we briefly describe the data reduction.
First, taking advantage of the large dithers of the images and of the
geometric distortion solution obtained by \citet{2022arXiv221203256G},
we extracted a grid of $5 \times 5$ library ePSFs both for the F115W
and F322W2 filters, following the procedure described by
\citet{2022MNRAS.517..484N}. By using the brightest and most isolated
stars in each image, we perturbed the ePSFs to take into account their
time variations and we used them to obtain positions and fluxes of the
sources detected in any given image (we refer to this as
\textit{first-pass} photometry).
For each filter, we transformed positions and magnitudes in a common
reference frame, defined by the Gaia\,DR3 catalogue
(\citealt{2021A&A...649A...1G}) and the first image in each filter. We
used images, perturbed ePSFs, and transformations to carry out the
so-called \textit{second-pass} photometry, by using the \texttt{KS2}
software, developed by J.\,Anderson \citep{2008AJ....135.2055A}, which
was employed and described in several works (e.g.,
\citealt{2017ApJ...842....6B, 2018MNRAS.481.3382N,
  2021MNRAS.505.3549S}), and here adapted to NIRCam images.
This routine, analysing all images simultaneously in a consistent
reference frame, allows us to go deeper than the first-pass photometry
and to detect extremely faint sources which would be otherwise lost in
the noise of individual exposures.
We carefully cleaned the final catalogue from the artifacts and bad
sources (in part due to the imperfection of the images) by using the
quality parameters output of the \texttt{KS2} routine
\citep{2008AJ....135.2055A}. Calibration of the instrumental
magnitudes were obtained as in \citet{2022MNRAS.517..484N}.

In this study, we made also use of the partially overlapped {\it HST}
Wide Field Camera 3 (WFC3) optical and IR observations collected during the
GO-11677 (PI: Richer, \citealt{2012AJ....143...11K}) in F606W, F110W,
and F160W. Astro-photometric catalogues were obtained from this
data-set following the data reduction procedure described by
\citet{2018MNRAS.481.3382N} and \citet{2018MNRAS.481.5339B,
  2020MNRAS.494.2068B}.

\begin{figure*}
\includegraphics[bb=19 366 531 717, width=0.9\textwidth]{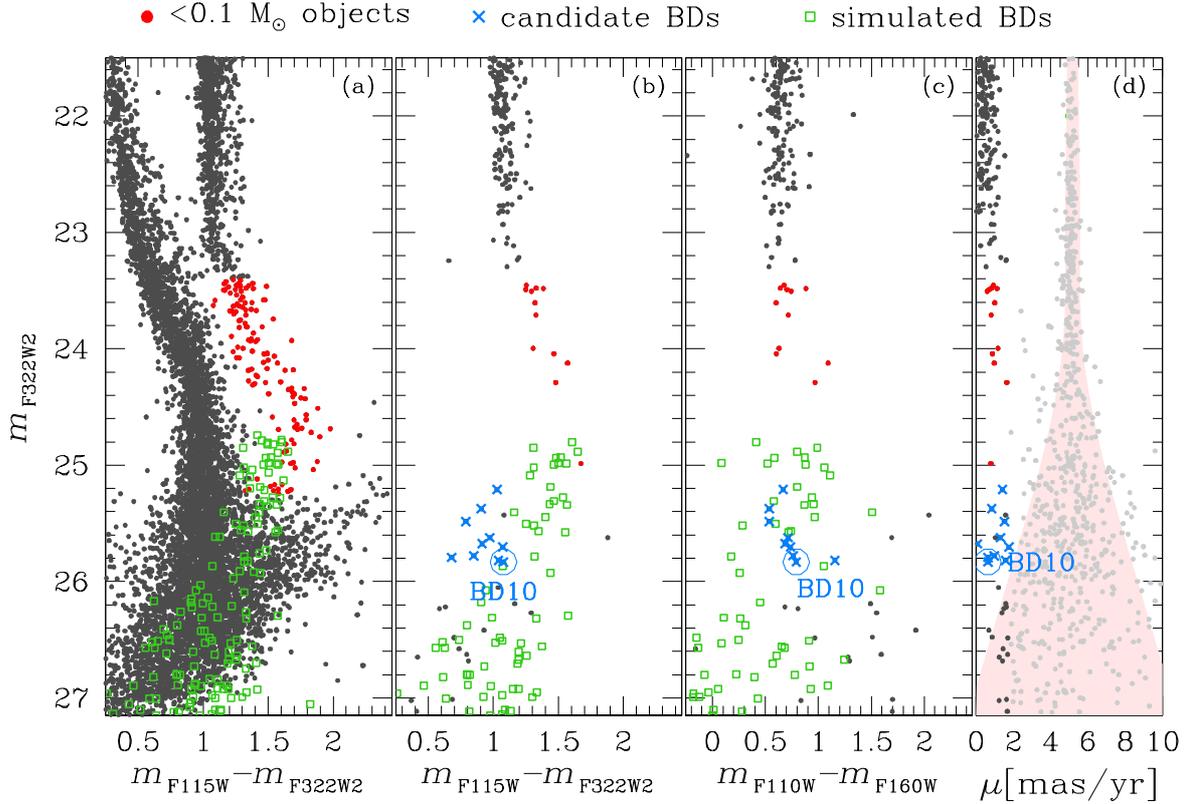}
\caption{Comparison between simulated BDs (green squares) and observed
  CMDs. In panel (a), all the stars measured in the {\it JWST} images
  are plotted in black, while in red are indicated objects consistent with 
  mass $<0.1~M_{\odot}$.  In next three panels to the left, for
  clearness purposes, we do not show those WDs clearly identified in
  Fig.~\ref{fig:1}. Panels (b) and (c) show the $m_{\rm F322W2}$
  versus $(m_{\rm F115W}-m_{\rm F322W2})$ and the $m_{\rm F322W2}$
  versus $(m_{\rm F110W}-m_{\rm F160W})$ CMDs, respectively, for the
  stars with proper motions in agreement with the mean cluster's
  motion. Azure crosses are the candidate BDs we identified in this
  study. Panel (d) shows the proper motion distribution referred to the
  mean cluster's motion. The shaded, rose region contains the stars
  whose proper motion is within $2\sigma$ from the mean motion of the
  SMC's stars. The azure circle is the faintest candidate BD we found,
  whose multi-filter finding charts is shown in Fig.~\ref{fig:bd10}. \label{fig:2}}
\end{figure*}

\begin{figure*}
\includegraphics[width=0.9\textwidth]{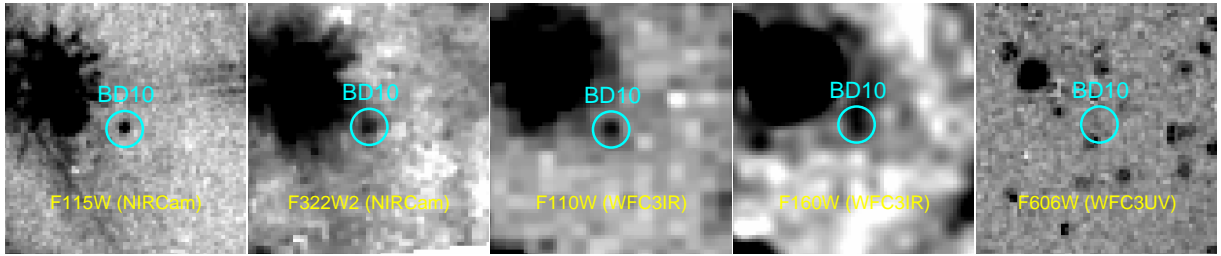}
\caption{Finding charts of the faintest candidate BD found in this
  paper. The star is visible in all the adopted IR bands,
  but not in the optical band F606W, as expected for a BD. \label{fig:bd10}}
\end{figure*}

\section{Colour-magnitude diagrams \& motions}
\label{sec:cmd}
In spite of the degradation suffered by these \textit{JWST} images,
our procedures were able to recover exquisite photometry and
astrometry, which allowed us to undertake an independent investigation
--on a quite different subject-- off the nominal main goal of program
GO-2560.
In this paper we present our study of the BDs of 47\,Tuc. 

In left-panel of Figure\,\ref{fig:0} we show the $m_{\rm F322W2}$
versus $m_{\rm F115W}-m_{\rm F322W2}$ CMD for the all stars in the
field, where the cluster's lower MS, the white dwarf (WD) sequence,
and the Small Magellanic Cloud (SMC) MS are all well populated and
clearly defined. Shaded regions show the 1, 3, 5\,$\sigma$ floor-noise
level (computed as described in \citealt{2023MNRAS.518.3722B}). We
overlapped (in green) a \texttt{BaSTI-IAC} 10\,Gyr isochrone
(\citealt{2018ApJ...856..125H,2021ApJ...908..102P}) to the MS of
47\,Tuc. We used the distance modulus $(m-M)_0=13.276$ tabulated by
\citet{2021MNRAS.505.5978V} and the reddening $E(B-V)\sim 0.04$ as
reported by \citet{1996AJ....112.1487H}.  Beside usual uncertainties
of models in reproducing the observed very low-mass MS stars, it is
also clear that the lower MS of 47\,Tuc (down to $m_{\rm F322W2}\sim
23$) is far from being well represented by a single stellar population
(as expected). A more detailed analysis of the multiple population
phenomenon in the lower MS of 47\,Tuc will be the subject of a future
article (Nardiello et al.\ in prep.). These theoretical models end at
the lower limit $M=0.1 M_\odot$ at magnitude $m_{\rm F322W2}\sim 23$.

In right-panel of Fig.\,\ref{fig:0} we highlighted in red the stars
that are likely cluster members and have masses $M<0.1~M_{\odot}$.
To interpret the sub-stellar objects we employed the BD tracks by
\citet[ATMO2020]{2020A&A...637A..38P} and computed 10\,Gyr isochrones
for solar-metallicity BDs (being not available to us BDs models at
47\,Tuc metallicities). The line in blue shows the isochrone obtained
from chemical equilibrium atmosphere (CEQ) models, while the one in
magenta shows the one obtained employing atmosphere with
non-equilibirum chemistry and ‘strong’ mixing (NEQ strong).
Although these models do not reflect the exact chemistry of the stars
in 47\,Tuc, it is possible to note that the most massive BDs
($M>0.07~M_{\odot}$) positions on the CMD are consistent with the
observed sources highlighted in red.

In order to verify the cluster membership of these low-mass stars
(indicated in red in Fig.\,\ref{fig:0}), and the membership of even
fainter stellar-object (which are completely lost in the MS of the SMC
in these CMDs), we reduced deep {\it HST} observations collected with
the WFC3 (see Sect.~\ref{sec:obs}).
These early epochs, along with the new \textit{JWST} positions, were
then employed to compute the relative proper motions (PMs) as in
\citet{2014MNRAS.439..354B}, by using 47\,Tuc's stars as reference
system for the motion, and MS stars as reference for the local
transformations. The average temporal baseline between the two epochs
is $\Delta t =12.4$~years. The resulting PMs are reported in panels
(c) of Fig.~\ref{fig:1}: we selected the likely cluster members by
tagging the stars inside the magenta circles as sources with a proper
motion in agreement with the cluster motion. Very red, cool objects
are not visible in the $m_{\rm F115W}$ versus $(m_{\rm F606W}-m_{\rm
  F115W})$ CMD, and we used this CMD (panel (a) of Fig.~\ref{fig:1})
to identify, among the likely cluster members, the WDs (in
blue) and other faint objects that can not be low MS stars/BDs (in
green). In panel (b) we show the stars for which we have proper
motions; likely WDs and other objects are marked as in panel
(a), while with a red circle we highlighted $<0.1 M_{\odot}$ stars and
candidate BDs.

In the next section we compare the observed CMDs with synthetic CMDs
obtained from theoretical models.

\section{Comparison with theoretical isochrones}

To obtain a qualitative estimate of the expected number of BDs at
different magnitudes and their approximate location in the CMD, we
first derived the mass-function (MF) for the observed MS stars,
employing the \texttt{BaSTI-IAC} models. The derived masses turned-out
to be rather flat in the mass range 0.15-0.35\,M$_\odot$.  We then
extrapolated this derived flat MF value down to the BDs mass range
(0.015-0.075\,M$_\odot$), and computed a synthetic CMD (ignoring
completeness) by employing the only BD isochrones at our disposal 
(described in previous section), which however are for Solar
metallicity.
To generate the synthetic CMD we proceeded as follows: we first
generated a random $\log_{10}(\rm Mass)$ in the interval
[$-1.8$,$-1.12$] following a uniform distribution.
For each random mass we associated a magnitude in the {\it JWST} and
{\it HST} filters using the BDs models, linearly interpolating between
the two nearest theoretical points.  We then added a random noise in
magnitude with errors as estimated from the real data.

Synthetic BDs are shown with green symbols (squares) in
Fig.\,\ref{fig:2}, while the observed sources with black points.
Panel (a) shows all the observed stars, while panels (b) and (c) only
the sub-sample of the observed stars with an estimate for the PMs, for
the CMDs $m_{\rm F322W2}$ versus $m_{\rm F115W}-m_{\rm F322W2}$
(\textit{JWST}-only) and $m_{\rm F322W2}$ versus $m_{\rm F110W}-m_{\rm
  F160W}$ (\textit{HST}-only), respectively.
Finally, panel (d) shows the combined 2-D proper motions as a function
of \textit{JWST}'s magnitude $m_{\rm F322W2}$.
This panel reveals an almost-perfect separation between field objects
(mainly SMC stars) and cluster members down to $m_{\rm F322W2} \simeq
26$, below which the membership become less clear.
We have identified (and highlighted with azure crosses) a group of ten
sources, which had proper motion consistent with the cluster's mean
motion, survived to all the selections and were located in the area of
the CMDs where BDs with mass of $\sim0.072 M_{\odot}$ are expected
(synthetic BDs in green).  We can also exclude that these objects are
foreground BDs, as those would be close-by, with considerably large
dispersion and high proper motions.\\~

In summary, in this observational effort, we have employed 
compromised public images collected with NIRCam at the
focus of \textit{JWST}, 
recovering exquisite
photometry and astrometry that enable scientific investigations.
We used our reduced data to explore the faintest stellar objects in a
field of 47\,Tuc, where we isolated a group of 10-objects that we
identify as candidate BD members of 47\,Tuc. Their membership is supported by
the following arguments:
\textit{(i)} their PMs are consistent with them being cluster members;  
\textit{(ii)} their location on CMD is qualitatively consistent with the expected location by approximate models; and finally, 
\textit{(iii)} the observed number of candidate BDs is consistent with the simulated number obtained extrapolating the MS MF into the BD domain, also considering all the uncertainties (models, MF extrapolated, completeness, etc.).

Only adequate \textit{JWST} follow-up observations will
be able to confirm the nature of these BD candidates as true members
of 47\,Tuc, by means of multi-wavelength deep observations and
\textit{JWST}-only derived proper-motion.

\section*{Acknowledgements}
The authors thank the anonymous referee for carefully reading the 
manuscript and for the useful suggestions.
DN, MG, and LRB acknowledge support by MIUR under PRIN
program \#2017Z2HSMF and by PRIN-INAF 2019 under program \#10-Bedin.
This work is based on observations made with the NASA/ESA/CSA {\it JWST}. 
The data were obtained from the Mikulski Archive
for Space Telescopes at the Space Telescope Science Institute, which
is operated by the Association of Universities for Research in
Astronomy, Inc., under NASA contract NAS 5-03127 for \textit{JWST}.
These observations are associated with program GO-2560 (PI: Marino).
This research is also based on observations made with the NASA/ESA
{\it Hubble Space Telescope} obtained from the Space Telescope Science
Institute, which is operated by the Association of Universities for
Research in Astronomy, Inc., under NASA contract NAS 5–26555. These
observations are associated with the program GO-11677 (PI: Richer).
This work has made use of data from the European Space Agency (ESA)
mission {\it Gaia} (\url{https://www.cosmos.esa.int/gaia}), processed
by the {\it Gaia} Data Processing and Analysis Consortium (DPAC,
\url{https://www.cosmos.esa.int/web/gaia/dpac/consortium}). Funding
for the DPAC has been provided by national institutions, in particular
the institutions participating in the {\it Gaia} Multilateral
Agreement.

\section*{Data Availability}
The data underlying this article are publicly available in the Mikulski
Archive for Space Telescopes at \texttt{https://mast.stsci.edu/}.
As electronic material, we provide coordinates and magnitudes,
for the 47\,Tuc's BD candidates identified in this work (see also Table~\ref{tab:0}). 

\begin{table*}
\caption{Candidate BDs identified in this work \label{tab:0}}
\begin{tabular}{l c c c c c c}
\hline
\hline
{\bf ID}  &  {\bf $\alpha$} & {\bf $\delta$}  &  {\bf $m_{\rm F115W}$}     &  {\bf $m_{\rm F322W2}$}   &  {\bf $m_{\rm F110W}$}   &  {\bf $m_{\rm F160W}$} \\
          &       (deg.)    &   (deg.)        &                          &                        &                        &                    \\
\hline
BD01   &  5.62526350   & $-$72.17662930    & 26.23      & 25.21  &     25.77  &     25.10 \\
BD02   &  5.63161207   & $-$72.15088600    & 26.27      & 25.48  &     26.10  &     25.56 \\
BD03   &  5.65459538   & $-$72.16556903    & 26.28      & 25.37  &     26.45  &     25.92 \\
BD04   &  5.60850256   & $-$72.16865613    & 26.47      & 25.79  &     26.29  &     --    \\
BD05   &  5.69237883   & $-$72.17938380    & 26.58      & 25.67  &     26.50  &     25.81 \\
BD06   &  5.60578561   & $-$72.15298701    & 26.59      & 25.62  &     26.68  &     25.96 \\
BD07   &  5.59451482   & $-$72.17008771    & 26.62      & 25.77  &     26.70  &     25.94 \\
BD08   &  5.58150990   & $-$72.16819902    & 26.77      & 25.70  &     27.56  &     26.82 \\
BD09   &  5.63124706   & $-$72.15303712    & 26.85      & 25.81  &     27.24  &     26.08 \\
BD10   &  5.62582295   & $-$72.15644952    & 26.90      & 25.83  &     26.82  &     26.03 \\
\hline
\end{tabular}
\end{table*}



\bibliographystyle{mnras}
\bibliography{biblio} 


\appendix

\section{Astro-ph bonus: compromised PSFs versus real PSFs}

\begin{figure*}
\includegraphics[width=0.4\textwidth]{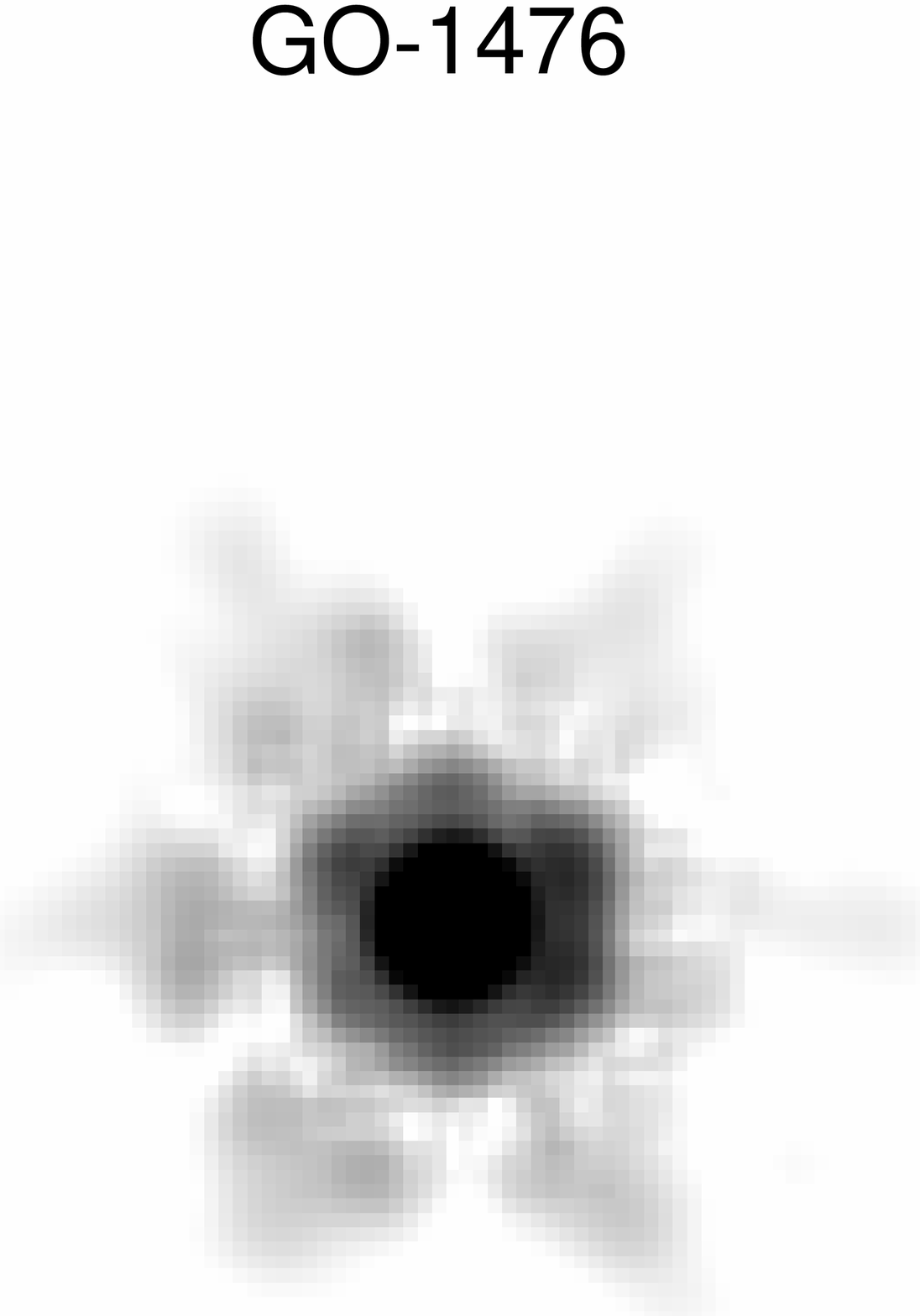}
\includegraphics[width=0.4\textwidth]{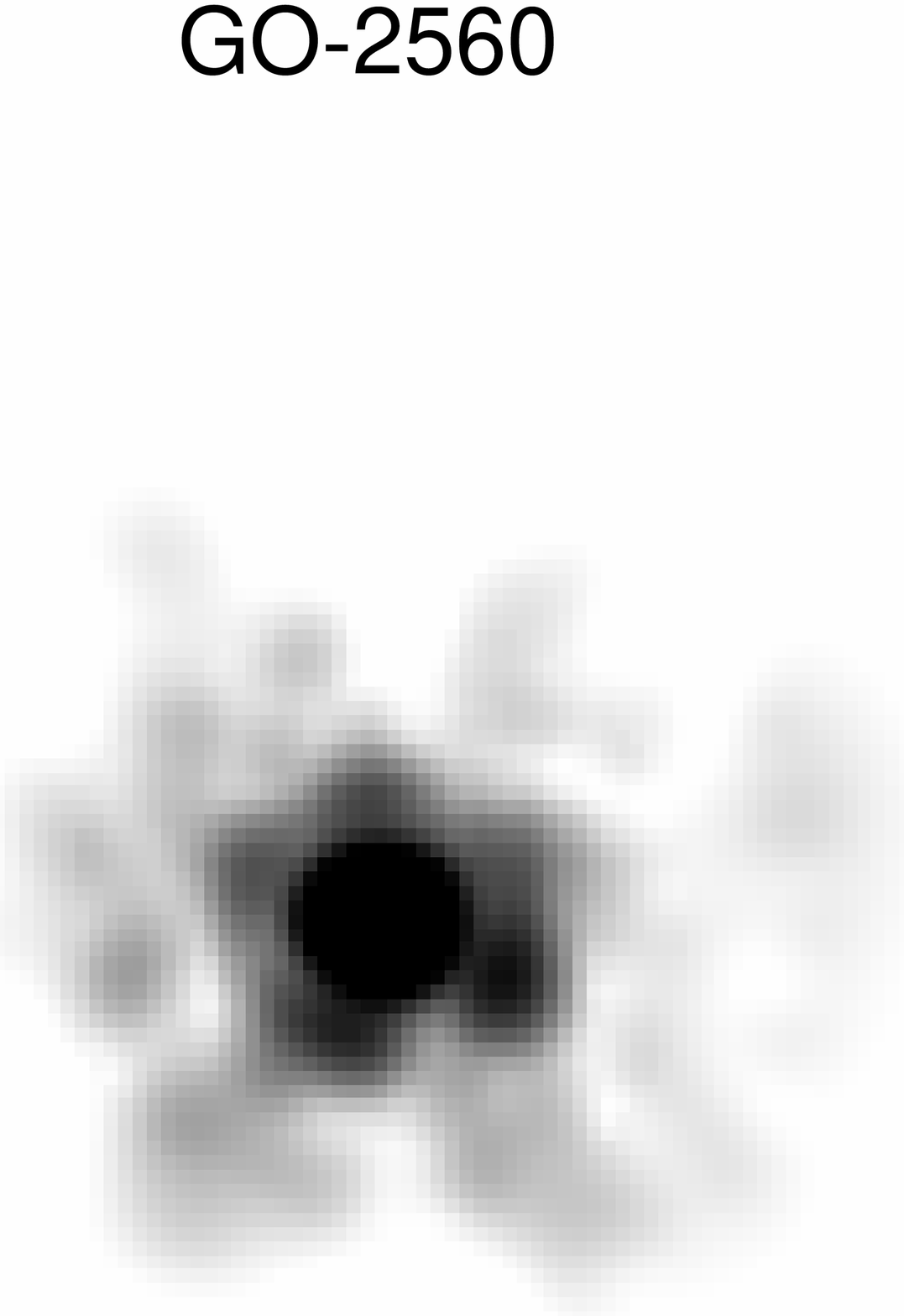}
\caption{Comparison between the same ePSF obtained from good data of
  GO-1476 (left) and from compromised images of GO-2560 (right). \label{fig:ap1}}
\end{figure*}

As reported in Sect.~\ref{sec:obs}, the data used in this work was
compromised by technical problems. These problems translate in stellar
shapes totally different from the real ones. In Fig.~\ref{fig:ap1} we
report the comparison between the same ePSF of the detector A1
obtained both with the compromised data from the GO-2560 and with the
good data of the Large Magellanic Cloud from GO-1476. The differences
between the two PSFs are clear.

\bsp	
\label{lastpage}
\end{document}